\documentstyle[12pt]{article}
\begin{document}
\newcommand {\be}{\begin{equation}}
\newcommand {\ee}{\end{equation}}
\newcommand {\bea}{\begin{array}}
\newcommand {\cl}{\centerline}
\newcommand {\eea}{\end{array}}
\renewcommand {\theequation}{\thesection.\arabic{equation}}
\renewcommand {\thefootnote}{\fnsymbol{footnote}}
\newcommand {\newsection}{\setcounter{equation}{0}\section}
\renewcommand {\thefootnote}{\fnsymbol{footnote}}
\renewcommand{\a}{\alpha}
\renewcommand{\b}{\beta}
\newcommand{\g}{\gamma}           \newcommand{\G}{\Gamma}
\renewcommand{\d}{\delta}         \newcommand{\D}{\Delta}
\newcommand{\ve}{\varepsilon}
\newcommand{\eps}{\epsilon}
\newcommand{\k}{\kappa}
\newcommand{\ld}{\lambda}        \newcommand{\LD}{\Lambda}
\newcommand{\om}{\omega}         \newcommand{\OM}{\Omega}
\newcommand{\p}{\psi}             \newcommand{\PS}{\Psi}
\newcommand{\ro}{\rho}
\newcommand{\s}{\sigma}           \renewcommand{\S}{\Sigma}
\newcommand{\th}{\theta}         \newcommand{\T}{\Theta}
\newcommand{\f}{{\phi}}           \newcommand{\F}{{\Phi}}
\newcommand{\vf}{{\varphi}}
\newcommand{\y}{{\upsilon}}       \newcommand{\Y}{{\Upsilon}}
\newcommand{\z}{\zeta}
\newcommand{\X}{\Xi}
\newcommand{\cA}{{\cal A}}
\newcommand{\cB}{{\cal B}}
\newcommand{\cC}{{\cal C}}
\newcommand{\cD}{{\cal D}}
\newcommand{\cE}{{\cal E}}
\newcommand{\cF}{{\cal F}}
\newcommand{\cG}{{\cal G}}
\newcommand{\cH}{{\cal H}}
\newcommand{\cI}{{\cal I}}
\newcommand{\cJ}{{\cal J}}
\newcommand{\cK}{{\cal K}}
\newcommand{\cL}{{\cal L}}
\newcommand{\cM}{{\cal M}}
\newcommand{\cN}{{\cal N}}
\newcommand{\cO}{{\cal O}}
\newcommand{\cP}{{\cal P}}
\newcommand{\cQ}{{\cal Q}}
\newcommand{\cS}{{\cal S}}
\newcommand{\cR}{{\cal R}}
\newcommand{\cT}{{\cal T}}
\newcommand{\cU}{{\cal U}}
\newcommand{\cV}{{\cal V}}
\newcommand{\cW}{{\cal W}}
\newcommand{\cX}{{\cal X}}
\newcommand{\cY}{{\cal Y}}
\newcommand{\cZ}{{\cal Z}}
\newcommand{\hA}{{\widehat A}}
\newcommand{\hB}{{\widehat B}}
\newcommand{\hC}{{\widehat C}}
\newcommand{\hD}{{\widehat D}}
\newcommand{\hE}{{\widehat E}}
\newcommand{\hF}{{\widehat F}}
\newcommand{\hG}{{\widehat G}}
\newcommand{\hH}{{\widehat H}}
\newcommand{\hI}{{\widehat I}}
\newcommand{\hJ}{{\widehat J}}
\newcommand{\hK}{{\widehat K}}
\newcommand{\hL}{{\widehat L}}
\newcommand{\hM}{{\widehat M}}
\newcommand{\hN}{{\widehat N}}
\newcommand{\hO}{{\widehat O}}
\newcommand{\hP}{{\widehat P}}
\newcommand{\hQ}{{\widehat Q}}
\newcommand{\hS}{{\widehat S}}
\newcommand{\hR}{{\widehat R}}
\newcommand{\hT}{{\widehat T}}
\newcommand{\hU}{{\widehat U}}
\newcommand{\hV}{{\widehat V}}
\newcommand{\hW}{{\widehat W}}
\newcommand{\hX}{{\widehat X}}
\newcommand{\hY}{{\widehat Y}}
\newcommand{\hZ}{{\widehat Z}}
\newcommand{\Ha}{{\widehat a}}
\newcommand{\Hb}{{\widehat b}}
\newcommand{\Hc}{{\widehat c}}
\newcommand{\Hd}{{\widehat d}}
\newcommand{\He}{{\widehat e}}
\newcommand{\Hf}{{\widehat f}}
\newcommand{\Hg}{{\widehat g}}
\newcommand{\Hh}{{\widehat h}}
\newcommand{\Hi}{{\widehat i}}
\newcommand{\Hj}{{\widehat j}}
\newcommand{\Hk}{{\widehat k}}
\newcommand{\Hl}{{\widehat l}}
\newcommand{\Hm}{{\widehat m}}
\newcommand{\Hn}{{\widehat n}}
\newcommand{\Ho}{{\widehat o}}
\newcommand{\Hp}{{\widehat p}}
\newcommand{\Hq}{{\widehat q}}
\newcommand{\Hs}{{\widehat s}}
\newcommand{\Hr}{{\widehat r}}
\newcommand{\Ht}{{\widehat t}}
\newcommand{\Hu}{{\widehat u}}
\newcommand{\Hv}{{\widehat v}}
\newcommand{\Hw}{{\widehat w}}
\newcommand{\Hx}{{\widehat x}}
\newcommand{\Hy}{{\widehat y}}
\newcommand{\Hz}{{\widehat z}}
\newcommand{\deff}{\,\stackrel{\rm def}{\equiv}\,}
\newcommand{\lra}{\longrightarrow}
\newcommand{\ra}{\,\rightarrow\,}
\def\limar#1#2{\,\raise0.3ex\hbox{$\longrightarrow$\kern-1.5em\raise-1.1ex
\hbox{$\scriptstyle{#1\rightarrow #2}$}}\,}
\def\limarr#1#2{\,\raise0.3ex\hbox{$\longrightarrow$\kern-1.5em\raise-1.3ex
\hbox{$\scriptstyle{#1\rightarrow #2}$}}\,}
\def\limlar#1#2{\ \raise0.3ex
\hbox{$-\hspace{-0.5em}-\hspace{-0.5em}-\hspace{-0.5em}
\longrightarrow$\kern-2.7em\raise-1.1ex
\hbox{$\scriptstyle{#1\rightarrow #2}$}}\ \ }
\newcommand{\limm}[2]{\lim_{\stackrel{\scriptstyle #1}{\scriptstyle #2}}}
\newcommand{\wt}{\widetilde}
\newcommand{\os}{{\otimes}}
\newcommand{\da}{{\dagger}}
\newcommand{\stimes}{\times\hspace{-1.1 em}\supset}
\def\h{\hbar}
\newcommand{\ih}{\frac{\i}{\h}}
\newcommand{\exx}[1]{\exp\left\{ {#1}\right\}}
\newcommand{\ord}[1]{\mbox{\boldmath{$\cO$}}\left({#1}\right)}
\newcommand{\one}{{\leavevmode{\rm 1\mkern -5.4mu I}}}
\newcommand{\Z}{Z\!\!\!Z}
%
\newcommand{\Ibb}[1]{ {\rm I\ifmmode\mkern
            -3.6mu\else\kern -.2em\fi#1}}
\newcommand{\ibb}[1]{\leavevmode\hbox{\kern.3em\vrule
     height 1.2ex depth -.3ex width .2pt\kern-.3em\rm#1}}
\newcommand{\N}{{\Ibb N}}
\newcommand{\C}{{\ibb C}}
\newcommand{\R}{{\Ibb R}}
\newcommand{\HH}{{\Ibb H}}
\newcommand{\rational}{{\kern .1em {\raise .47ex
\hbox{$\scripscriptstyle |$}}
    \kern -.35em {\rm Q}}}
\newcommand{\bm}[1]{\mbox{\boldmath${#1}$}}
\newcommand{\intf}{\int_{-\infty}^{\infty}\,}
\newcommand{\LL}{\cL^2(\R^2)}
\newcommand{\LLS}{\cL^2(S)}
\newcommand{\Ree}{{\cal R}\!e \,}
\newcommand{\Imm}{{\cal I}\!m \,}
\newcommand{\tr}{{\rm {Tr} \,}}
\newcommand{\er}{{\rm{e}}}
\renewcommand{\i}{{\rm{i}}}
\newcommand{\divv}{{\rm {div} \,}}
\newcommand{\id}{{\rm{id}\,}}
\newcommand{\ad}{{\rm{ad}\,}}
\newcommand{\Ad}{{\rm{Ad}\,}}
\newcommand{\const}{{\rm{\,const\,}}}
\newcommand{\rank}{{\rm{\,rank\,}}}
\newcommand{\diag}{{\rm{\,diag\,}}}
\newcommand{\sign}{{\rm{\,sign\,}}}
\newcommand{\pa}{\partial}
\newcommand{\pad}[2]{{\frac{\partial #1}{\partial #2}}}
\newcommand{\padd}[2]{{\frac{\partial^2 #1}{\partial {#2}^2}}}
\newcommand{\paddd}[3]{{\frac{\partial^2 #1}{\partial {#2}\partial {#3}}}}
\newcommand{\der}[2]{{\frac{{\rm d} #1}{{\rm d} #2}}}
\newcommand{\derr}[2]{{\frac{{\rm d}^2 #1}{{\rm d} {#2}^2}}}
\newcommand{\fud}[2]{{\frac{\delta #1}{\delta #2}}}
\newcommand{\fudd}[2]{{\frac{\d^2 #1}{\d {#2}^2}}}
\newcommand{\fuddd}[3]{{\frac{\d^2 #1}{\d {#2}\d {#3}}}}
\newcommand{\dpad}[2]{{\displaystyle{\frac{\partial #1}{\partial #2}}}}
\newcommand{\dfud}[2]{{\displaystyle{\frac{\delta #1}{\delta #2}}}}
\newcommand{\dd}{\partial^{(\ve)}}
\newcommand{\ddd}{\bar{\partial}^{(\ve)}}
\newcommand{\dfrac}[2]{{\displaystyle{\frac{#1}{#2}}}}
\newcommand{\dsum}[2]{\displaystyle{\sum_{#1}^{#2}}}
\newcommand{\dint}{\displaystyle{\int}}
\newcommand{\dg}{\!\not\!\partial}
\newcommand{\vg}[1]{\!\not\!#1}
\def\<{\langle}
\def\>{\rangle}
\def\lgl{\langle\langle}
\def\rgr{\rangle\rangle}
\newcommand{\bra}[1]{\left\langle {#1}\right|}
\newcommand{\ket}[1]{\left| {#1}\right\rangle}
\newcommand{\vev}[1]{\left\langle {#1}\right\rangle}
\newcommand{\bn}{\begin{eqnarray}}
\newcommand{\en}{\end{eqnarray}}
\newcommand{\bnn}{\begin{eqnarray*}}
\newcommand{\enn}{\end{eqnarray*}}
\newcommand{\e}{\label}
\newcommand{\nbr}{\nonumber\\[2mm]}
\newcommand{\r}[1]{(\ref{#1})}
\newcommand{\refp}[1]{\ref{#1}, page~\pageref{#1}}
\newcommand{\qq}{\qquad}
\newcommand{\qqq}{\quad\quad}
\newcommand{\biz}{\begin{itemize}}
\newcommand{\eiz}{\end{itemize}}
\newcommand{\ben}{\begin{enumerate}}
\newcommand{\een}{\end{enumerate}}
\def\nc{noncommutative }
\baselineskip 0.65 cm
\begin{flushright}
hep-th/0101209\\
{\bf HIP-2001-01/TH}\\
IC/2001/3 \\
\today
\end{flushright}
\begin{center}
\vspace*{1.0cm}

{\Large \bf Quantum Theories on Noncommutative Spaces\\
with Nontrivial Topology:\\ \vspace{7mm}Aharonov-Bohm and Casimir
Effects }

\vskip 1cm

{\large {\bf M. Chaichian}}$^{\dagger}$,
\ \ {\large{\bf
A. Demichev}}$^{\dagger,}$\renewcommand{\thefootnote}{a}
\footnote{Permanent address:
Nuclear Physics Institute, Moscow State University,
119899 Moscow, Russia},
\ \ {\large{\bf
P. Pre\v{s}najder}}$^{\dagger,}$\renewcommand{\thefootnote}{b}
\footnote{Permanent address:
Department of Theoretical Physics, Comenius University,
Mlynsk\'{a} dolina, SK-84248 Bratislava, Slovakia},
\vskip 0.2cm
\ \ {\large {\bf M. M. Sheikh-Jabbari}}$^{\dagger\dagger}$
\ \ and \ \ {\large{\bf A. Tureanu}}$^{\dagger}$
\vskip 0.2cm

$^{\dagger}$High Energy Physics Division, Department of Physics,\\
University of Helsinki\\
\ \ {\small\it and}\\
\ \ Helsinki Institute of Physics,\\
P.O. Box 9, FIN-00014 Helsinki, Finland\\
\vskip 0.2cm
$^{\dagger\dagger}$ The Abdus Salam International Center for Theoretical Physics,\\
Strada Costiera 11,Trieste, Italy

\end{center}

\setcounter{footnote}{0}

\vspace{0.5 cm}
\begin{abstract}

After discussing the peculiarities of quantum systems on noncommutative (NC) spaces with nontrivial topology
and the operator representation of the $\star$-product on them, we consider the Aharonov-Bohm and Casimir
effects for  such spaces. For the case of the Aharonov-Bohm effect, we have obtained an explicit expression
for  the shift of the phase, which is  gauge invariant in the NC sense. The Casimir energy of a field theory
on a NC cylinder is divergent, but it becomes finite on a torus, when the dimensionless parameter of
noncommutativity is a rational number. The latter corresponds to a well-defined physical picture. Certain
distinctions from other treatments based on a different way of taking the noncommutativity into account are
also discussed.

\end{abstract}
\newpage
\newsection{Introduction}
The noncommutative geometry was formulated originally on
noncommutative analogs of Euclidean spaces, without a
distinguished notion of a physical time; a survey can be found in
the books \cite{Co}-\cite{GB}. The Euclidean field theories on a
\nc sphere (and also plane and cylinder) have been investigated in
refs. \cite{GKP1}.

However, from the point of view of physical applications, it is
desirable to consider models with a physical time. The \nc analog
of a Minkowski plane was originally introduced in \cite{Sny} and
investigated in the context of \nc geometry in \cite{DFR}. In this
approach, the space-time coordinates satisfy non-trivial
commutation relations:
\be\label{cr}
[\hat{x}_{\mu},\hat{x}_{\nu}]=\i\theta_{\mu\nu}\ ,
\ee
were $\theta_{\mu\nu}$ is a constant anti-symmetric tensor, in a suitable representation. 
Recently, it was found
that (\ref{cr}) follows naturally as a particular low-energy limit
of string theories \cite{SW}-\cite{ArJa1}, with $\theta_{\mu\nu}$
directly related to a constant antisymmetric background field
$B_{\mu\nu}$ in the presence of a D-brane. The explicit presence of the constant $\theta_{\mu\nu}$
in (\ref{cr}) violates Lorentz invariance (if the dimension of
space-time is greater than two).

Later, it was found that in such \nc Minkowski spaces, the ultraviolate divergences of 
quantum field theory (QFT) persist \cite{Filk},\cite{CDP1}.
Moreover, if the time is noncommutative, such field theories violate
unitarity and causality \cite{Se}-\cite{CDP3}.

On the contrary, models on a \nc space but with a commutative
time do not encounter such difficulties. Similarly, no principal
problems arise in quantum mechanics defined on \nc spaces with a
standard (commutative) time evolution. Some phenomenological consequences within such an approach for the Lamb
shift have been recently calculated (see \cite{Lamb} and refs. therein).

Recently it was discussed that the effective \nc field theories, in particular \nc Chern-Simons theory,
can serve as a natural description for the (fractional) quantum Hall effect \cite{NCCS}.
This may shed more light on the Wigner crystal-quantum Hall fluid phase transition.
Then the \nc Aharonov-Bohm, being a 2+1 dimensional effect on the \nc plane (or punctured plane), is of
great importance. In fact the Aharonov-Bohm phase is the phase which appears in front of the
wave-function of two charged particles upon their exchange (for a review, see
\cite{Chern}). Here we try to study this
problem, both semi-classically (and in first order in $\theta$, where $\theta$ is the dimensionful scale of
the tensor $\theta_{\mu\nu}$) and analytically.

The \nc models specified by (\ref{cr}) can be realized in terms
of a $\star$-product: the commutative algebra ${\cA}_0$ of functions
with the usual product $(fg)(x)=f(x)g(x)$ is replaced by the
$\star$-product Moyal algebra:
\begin{eqnarray}\label{star}
(f*g)(x)&=&\exx{{\i\over 2}\theta_{\mu\nu}
\partial_{x_{\mu}}\partial_{y_{\nu}}}f(x)g(y)\Big|_{x=y}\nbr
&=& f(x)g(x)+{\i\over2}\{f,g\}(x)+\ord{\theta^2}\ ,
\end{eqnarray}
where $\{f,g\}=\theta_{\mu\nu}(\partial_{\mu}f)(\partial_{\nu}g)$
is the Poisson bracket associated with $\theta_{\mu\nu}$. Such
associative $\star$-products have been proved to exist as a formal
power series for any Poisson bracket
$\{f,g\}=\theta_{\mu\nu}(x)(\partial_{\mu}f)(\partial_{\nu}g)$,
with a most general $x$-dependent $\theta_{\mu\nu}(x)$
\cite{Ko}. However, in general, the problems of the summability
and unitarizability (realization in terms of operators in a
Hilbert space) remain open.

The point is that the formal power series expansion (\ref{star})
does not take into account the global topological properties
and/or boundary conditions which are essential for the operator
realization. Here, we shall analyze four simple cases: the plane
${\R}^2$, the cylinder $C={\R}\times S^1$, the torus
$T^2=S^1\times S^1$ and the punctured plane
${\R}_{0}^2={\R}^2\setminus\{0\}$. They all are related to the
same Poisson bracket $\{f,g\}$ generated from the elementary
bracket $\{x_1,x_2\}=1$.

In section  2, we investigate the Aharonov-Bohm effect, first on the whole NC-plane and then on a punctured
plane, together with a non-trivial problem of implementation of the gauge invariance. The punctured plane is
topologically equivalent to the cylinder, but geometrically different from it. The next two sections 3 and 4
are devoted to the Casimir energies for a scalar field theory on a \nc cylinder and torus, respectively. The
usual Casimir effect concerns the vacuum energy between two plates (lines, in our 2D setting). However, it is
a delicate problem to introduce well-defined lines in \nc spaces with Dirichlet or Neumann boundary
conditions. We shall not discuss such \nc effects in the present paper. Section 5 contains conclusions and
discussion.

\newsection {Aharonov-Bohm effect on a noncommutative plane\e{BAep}}

The Aharonov-Bohm effect concerns the shift of the interference pattern in the double-slit experiment, due to
the presence of a thin long solenoid placed just between the two slits
\cite{Feynman,Holstein}.
Although the magnetic field $B$  is present only inside the solenoid, the corresponding
Schroedinger equation
depends explicitly on the magnetic potential $A$ (non-vanishing outside the solenoid). Therefore, the wave
function depends on $A$ and consequently the interference pattern shifts. However, due to the gauge
invariance, the shift in the phase of the particles propagator, $\delta\phi_0$, is gauge
invariant itself and can be expressed in non-local
terms of $B$. In the quasi-classical approximation, $\delta\phi_0={e\over 2 \pi\hbar c}\Phi$,
where $\Phi=B\pi\rho^2$ is the magnetic flux through
the solenoid of radius $\rho$. This effect has been confirmed experimentally \cite{Tono}.

In this section,  first we present the quasi-classical approach to the Aharonov-Bohm effect on a
NC-plane for a thin, but of finite radius, solenoid. In particular, we give the modifications to the phase shift,
$\delta\phi_0$, due to noncommutativity up to the first order in $\theta$. A short version of this part was
reported previously in \cite{ABPRL}.
Then in the next subsection using the algebraic method, we analyze the
effect for an infinitesimally thin solenoid, on a punctured NC-plane, for all orders in $\theta$.

\subsection{Path integral approach and quasi-classical approximation}

In this section, we shall first describe the Moyal
$\star$-product on a two dimensional plane ${\R}^2$. On ${\R}^2$, a Poisson bracket can be generated from the
elementary bracket
\be\label{Poissonbr}
\{x_1,x_2\}=1\ .
\ee
The commutative algebra
${\cal A}_0$ of functions on ${\R}^2$ is formed by functions of the
form:
\be\label{expan} 
f(x)={1\over 2\pi}\int
d^{2}k\,\tilde{f}(k)\er^{\i kx}\ ,\qq kx=k_1 x_1+k_2 x_2\ . 
\ee
Then, the Moyal product can be expressed as
\be\label{starpr}
(f\star g)(x)={1\over (2 \pi)^2}\int d^2
k_1d^2k_2\tilde{f}(k_1)\tilde {g}(k_2)\er^{-{\i\over 2}
\theta_{\mu\nu}k_1^{\mu}k_2^{\nu}}\er^{\i(k_1+k_2)x}\ , \ee
where
$\theta_{\mu\nu}=\theta\epsilon_{\mu\nu},\ \theta$-constant,
$\epsilon_{\mu\nu}$-anti-symmetric. This defines the corresponding
\nc algebra of functions $\cal A$ on ${\R}^2$.

In $\cA$, we can introduce the scalar product as:
\begin{eqnarray}
(f,g)&=&\int d^2x\bar{f}(x)\star g(x)=\int d^2x\bar{f}(x)g(x)\cr
&=&\int d^2k\bar{\tilde{f}}(k)\tilde{g}(k)\ .
\end{eqnarray}
Here, we have used the well-known fact that in the integrals containing as integrand a $\star$-product of two
functions, their $\star$-product can be replaced by a standard one.

Alternatively, one can start from an operator algebra generated by
the hermitian operators $\hat{x}_1$ and $\hat{x}_2$, satisfying
the commutation relation \be\label{comrel}
[\hat{x}_1,\hat{x}_2]=\i\theta. \ee The corresponding \nc algebra
$\cal A$ can be given as the algebra of operators of the form
\be\label{opalg} f(\hat {x})={1\over 2\pi} \int
d^{2}k\,\tilde{f}(k)\er^{\i k\hat{x}}\ ,\qq
k\hat{x}=k_1\hat{x}_1+k_2\hat{x}_2\ . \ee It can be seen easily
that the product in the operator algebra (\ref{opalg}) possesses
an expansion in powers of $\theta$, exactly corresponding to the
Moyal product.

The Hilbert space $\cH$ of quantum mechanics on a \nc plane is formed by the normalizable functions
$\Psi(x)\in \cA$, with finite norm. The wave function is an element from $\cH$, normalized to unity.
We should remind that in the \nc case, the usual physical meaning of wave functions as probability
amplitudes fails, and wave functions are just symbols \cite{CDP1}.
The operators $P_i$ and $Q_i$ acting in $\cH$ and satisfying Heisenberg canonical commutation relations are
defined by:
\be\label{op}
P_i\Psi(x)=-\i\partial_i\Psi(x)\ ,\qq X_i\Psi(x)=x_i\star \Psi(x)\ .
\ee

The problem of a particle moving in an external magnetic field on a \nc plane is specified by the Hamiltonian:
\be
H={1\over 2}(P_i+A_i)_{\star}^2={1\over 2}(P_i+A_i)\star (P_i+A_i)\ .
\ee
We now define the \nc analog of gauge transformations by
\footnote{In the present paper, our exponentials are defined by the Taylor series with the
$\star$-product, i.e. $\er^{i\lambda(x)}=1+i\lambda-{1\over 2}\lambda\star\lambda+\cdots$.}:
\begin{eqnarray}\label{gaugetransf}
\Psi(x)&\rightarrow &\er^{\i\lambda(x)}\star\Psi(x)\ ,\qq \lambda(x)-real\cr
A_i(x)&\rightarrow&\er^{\i\lambda(x)}\star A_i(x)\star
\er^{-\i\lambda(x)}
-\i\er^{\i\lambda(x)}\star(\partial_i\er^{-\i\lambda(x)})\ .
\end{eqnarray}
We point out the non-Abelian character of (\ref{gaugetransf}), due to the noncommutativity of the plane.
Consequently, the field strength is given by a non-Abelian formula, too:
\be
F(x)=\epsilon_{ij}(\partial_{i}A_{j}(x)+A_{i}(x)\star A_{j}(x))\ .
\ee
One can easily see that
\be\label{trgauge}
P_i+A_i\rightarrow\er^{\i\lambda(x)}\star(P_i+A_i)\star\er^{-\i\lambda(x)}\ ,
\ee
just like in a usual commutative case. Hence, the transition amplitude $(\Psi_f,\er^{-\i H t}\Psi_i)$ is gauge
invariant:
\be
(\Psi_f,\er^{-\i H t}\Psi_i)\rightarrow (\er^{\i\lambda}\star\Psi_f,\er^{\i\lambda}\star\er^{-\i H
t}\star\er^{-\i\lambda}\star\er^{\i\lambda}\star\Psi_i)=(\Psi_f,\er^{-\i H t}\Psi_i)\ .
\ee

In quantum mechanics, the exponents of the operators (e.g., $\er^{-\i H t}$) often do not correspond to local
operators. However, they can be conveniently represented by bi-local kernels. This is true in the \nc frame,
also. It can be easily seen that to any operator $K=K(P_i,X_i)=K(-\i\partial_i,x_i\star)$, cf. (\ref{op}), we can assign a kernel (a bi-local symbol)
$\cK(x,y)\in\cA\otimes\cA$, defined by:
\be
\cK(x,y)={1\over 2\pi}\int d^2q(K\er^{\i q x})\er^{-\i q y}
\ee
(we omit the symbol $\otimes$ for the direct product). The action of $K$ in terms of the kernel is
\be\label{action.K}
(K\Phi)(x)=\int d^2y\cK(x,y)\star\Phi(y)=\int d^2y\cK(x,y)\Phi(y)\ .
\ee
where we have used the fact that the $\star$-product in the
quadratic terms under the integral can be removed.
Thus, the matrix elements of $K$ are given as
\be
(\Psi,K\Phi)=\int d^2x d^2y\bar{\Psi}(x)\cK(x,y)\Phi(y)\ .
\ee
For a product of two operators, one can use the standard formula for the kernel composition
\begin{eqnarray}\label{prodop1}
(\cG\cK)(x,y)&=&\int d^2z\cG(x,z)\star\cK(z,y)\cr
&=&\int d^2z\cG(x,z)\cK(z,y)\ .
\end{eqnarray}
 Alternatively, one can use the formula
\be\label{prodop2}
(\cG\cK)(x,y)=\int{d^2q\over 2\pi}(G\er^{\i q x})\overline{(K^{\dagger}\er^{\i q y})}\ .
\ee
The proof of (\ref{prodop2}) is straightforward.

The kernel corresponding to the operator $\er^{-\i H t}$ will be denoted by $\cK_t(x,y)$ and called
propagator:
\be
\cK_t(x,y)=\int{d^2q\over 2\pi}(\er^{-\i H t}\er^{\i q x})\er^{-\i q y}\ .
\ee
{}From the product formula (\ref{prodop1}) and the identity $\er^{-\i H t_1}\er^{-\i H t_2}=\er^{-\i
H (t_1+t_2)}$, the usual composition law follows:
$$
\cK_{t_1+t_2}(x,y)=\int d^2z\cK_{t_1}(x,z)\cK_{t_{2}}(z,y)\ .
$$
Iterating this formula $N$ times and taking the limit $N\rightarrow\infty$, 
by standard arguments 
we arrive at the path integral representation of the propagator:
\be\label{propagator}
\cK_t(x,y)=\lim_{N\rightarrow\infty}\int d^2x_{N-1}\cdots d^2x_1\cK_{\epsilon}(x,x_{N-1})\cdots\cK_{\epsilon}(x_2,x_1)\cK_{\epsilon}(x_1,y)\ ,
\ee
with $\epsilon=t/N$. We stress that, due to the iterative procedure used to derive (\ref{propagator}), 
there is no need to use $\star$-product between the $\cK_{\epsilon}$'s, since in each step the $\star$-product
can be removed (see (\ref{action.K})).

The formula for the gauge transformation of the propagator follows directly from  eq. (\ref{trgauge}). In
fact, (\ref{trgauge}) implies:
\be
\er^{-\i H t}\rightarrow\er^{-\i \lambda}\star\er^{-\i H t}\star\er^{\i \lambda}
\ee
(as operators), so that
\begin{eqnarray}
\cK_t(x,y)&\rightarrow&(\er^{\i \lambda}\star \er^{-\i H t}\star \er^{-\i \lambda})(x,y)\cr
&=&\int dq(\er^{\i \lambda}\er^{-\i H t}\er^{\i q x})\overline{\er^{\i \lambda}\er^{\i q y}}\cr
&=&\int dq\er^{\i \lambda(x)}\star(\er^{-\i H t}\er^{\i q x})\overline{\er^{\i \lambda(y)}\star\er^{\i q y}}\cr
&=&\er^{\i \lambda(x)}\star\cK_t(x,y)\star\er^{-\i \lambda(y)}\ .
\end{eqnarray}
This is exactly the expected formula (here, the $\star$-product cannot be omitted).

As the next step, we shall calculate the short-time propagator $\cK_{\epsilon}(x,y)$ entering
(\ref{propagator}) to the first orders in $\epsilon$ and $\theta$. Using
\begin{eqnarray}
\cK_{\epsilon}(x,y)&=&\int{d^2p\over 2\pi}[1-{\i\epsilon\over 2}(P_i+A_i)^2+\cdots]\er^{\i p x}\er^{-\i p y}\cr
&=&\int{d^2p\over 2\pi}[\er^{\i p x}\er^{-\i p y}-{\i\epsilon\over 2}(P_i+A_i)\er^{\i p x}\overline{(P_i+A_i)\er^{\i p
y}}+\cdots]\cr
&=&\int{d^2p\over 2\pi}[\er^{\i p x}\er^{-\i p y}-{\i\epsilon\over 2}(p_i\er^{\i p x}+A_i(x)\star\er^{\i p
x})\overline{(p_i\er^{\i p y}+A_i(y)\star\er^{\i p y})}+\cdots]\ 
\end{eqnarray}
we write this in the form:
\be
\cK_{\epsilon}(x,y)=\int{d^2p\over 2\pi}\er^{\i p (x-y)-{\i\epsilon}H_e(p,\bar{x})}\ ,\qq
\bar{x}={1\over 2}(x+y)\ ,
\ee
where the effective Hamiltonian, $H_e$, is given as:
\be
H_e\cong{1\over 2}(\Pi_i+A_i(\bar{x}))^2\ ,\qq \Pi_i=p_i-{1\over
2}\theta_{jk}(\partial_{j}A_i(\bar{x}))p_{k}\ .
\ee
The symbol $\cong$ means equality in the first order in $\epsilon$ and $\theta$. Performing the $d^2p$
integration, we obtain the effective Lagrangian:
\be\label{efflagr}
\cL\cong{1\over 2}V_i V_i-V_i A_i(\bar{x})\, \qq V_i=v_i+{1\over
2}\theta_{jk}\partial_{j}A_i(\bar{x})v_{k}\ .
\ee
The formula for $\cK_{\epsilon}(x,y)$ then reads:
$\ \
\cK_{\epsilon}(x,y)\cong\er^{\i\int dt\cL(\bar{x}(t),\dot{\bar{x}}(t))}\ ,
$
where the effective action is calculated for a linear path, starting at $x_i(0)=x_i$ and terminating at
$x_i(\epsilon)=y_i$, i.e., $v_i=(y_i-x_i)/\epsilon$ and $A_i(\bar{x})=A_i({x+y\over 2})$. Up to terms linear
in $\theta$, the Lagrangian, with all physical constants included, becomes:
\be\label{lagr}
\cL={m\over 2}\vec{v}^2-{e\over c}\vec{v}\cdot\vec{A}-{e m\over 4\hbar
c}\vec{\theta}\cdot\bigl[v_i(\vec{v}\times\vec{\nabla}A_i)-
{e\over mc}v_i(\vec{A}\times\vec{\nabla}A_i)\bigr],   
\ee
where $\vec\theta=\theta\  \hat{z}$ and $\hat{z}$ is the unit vector normal to the $(x_1,\ x_2)$
plane and the cross product is the usual three dimensional one.
Thus, the total shift of phase for the Aharonov-Bohm effect, including the contribution due to
noncommutativity, will be:
\be\label{ncshift}
\delta\phi_{total}=\delta\phi_0+\delta\phi_{\theta}^{NC}\ ,
\ee
where $\delta\phi_0={e\over \hbar c}\oint d\vec{r}\cdot\vec{A}=
{e\over \hbar c}\int \vec{B}\cdot d\vec{S}={e\over \hbar c}\Phi$ ($\Phi$ being
the magnetic flux through the surface bounded by the closed path)
is the usual (commutative) phase shift and
\be\label{nccorr}
\delta\phi_{\theta}^{NC}={e m\over 4\hbar^2 c}\vec{\theta}\cdot
\oint dx_i\bigl[(\vec{v}\times\vec{\nabla}A_i)-
{e\over mc}(\vec{A}\times\vec{\nabla}A_i)\bigr]
\ee       
represents the \nc corrections.
For a finite-radius solenoid, the vector potential $\vec{A}$ entering (\ref{lagr})-(\ref{nccorr}) is given by:
\be\label{solenoid}
\vec{A}={1\over 2}B{\rho^2\over r}\vec{n}\ ,\qq r>\rho\ ,
\ee
where $B$ is the constant magnetic field inside the solenoid, $\rho$ is the radius of the solenoid and $\vec{n}$
is the unit vector orthogonal to $\vec{r}$.

The expression for the correction $\delta\phi_{\theta}^{NC}$ to the usual Aharonov-Bohm phase due to
noncommutativity can be explicitly obtained from (\ref{nccorr}) and (\ref{solenoid}). In an analogous way as in
the usual Aharonov-Bohm case \cite{Feynman,Holstein}, the calculation can be done by taking the closed classical
path (what is valid according to the experimental setup), which starts from the source and reaches the point
on the screen by passing through one of the two slits and returns to the source point through the other slit.

\subsection{Exact treatment of the Aharonov-Bohm effect on a punctured plane}

In the following, we describe the \nc version of this quantum mechanical
problem on a punctured plane ${\R}_{0}^{2}$, specified by the
following Hamiltonian:
\be\label{ham}
\hH={1\over 2}(-\i\partial_{j}+A_{j}^s)^{2}\ ,\qq
A_{j}^s=\mu\epsilon_{jk}\frac{x_k}{r^2}\ ,
\ee
with $x=(x_1,x_2)\in {\R}^2_{0}$ and $r^2=x^2_{1}+x_2^{2}$; $A^s_j$ represents
the magnetic field generated by a thin solenoid located at the
origin, perpendicular to the plane. The Hamiltonian itself acts on
a suitable domain in the Hilbert space $\cH=L^{2}({\R}^2,d^{2}x)$
with the scalar product
\be\label{scprod}
(\Psi,\Phi)=\int d^{2}x\bar{\Psi}(x)\Phi(x)\ .
\ee

It is worth noting that the punctured plane is a relevant one in the physical situations. The
reason is that, in the case of the solenoid (and also in the exchange phase problem in the anyonic
system \cite{Chern}), the location of the thin solenoid (and the anyonic magnetic flux) should
be excluded from the physically available space.

In order to define a \nc generalization of the configuration
space ${\R}_{0}^{2}$, one can not start with an {\it ad hoc} Poisson
structure generated by:
\be\label{Poisson}
\{x_1,x_2\}_0=1\ ,\qq (x_1,x_2)\in{\R}_{0}^{2}\ ,
\ee
since the Darboux pair $(x_1,x_2)$ does not take into account
the topological properties of ${\R}_{0}^{2}$.

Therefore, we map ${\R}_{0}^{2}$ on a cylinder $C={\R}\times S^1$: $\eta={1\over
2}\ln{(x_{i}^2/\rho^2)}$, $\phi={1\over 2} \arctan{(x_{1}/x_{2})}$
($\rho$ is a constant of the dimension [length]
which will be put equal to 1 below). The cylinder (as a $E(2)$ co-adjoint
orbit) possesses the natural Poisson bracket given by $\{\eta,\phi\}=1$, or
equivalently by
\be\label{Poissonnew}
\{\eta,U\}=iU\ ,\qq \eta\in{\R},\ U=\er^{\i  \phi}\in S^1\ .\ee
This
generates the Poisson structure in the space of  periodic functions in the
polar angle $\phi$. In what follows, we shall use this Poisson structure and
not the one given by (\ref{Poisson}).

Let us rewrite first the commutative version of our problem in the variables
$\{\eta,\phi\}=1$. Expanding the wave functions as
\be
\Psi=\sum_{k\in\Z}a_k(\eta)\er^{\i k \phi}\ ,\qq
\Phi=\sum_{k\in\Z}b_k(\eta)\er^{\i k \phi}\ ,
\ee
the scalar product in $\cH=L^2(\R_0^2,\er^{2\eta}d\eta d\phi)$ reads:

\be\label{scprodnew}
(\Psi,\Phi)=\int d\eta d\phi\,
\er^{2\eta}\,\bar{\Psi}(\eta,\phi)\Phi(\eta,\phi)=2\pi \sum_{k} \int_{{\R}}d\eta\,
\er^{2\eta}\bar{a}_{k}(\eta)b_{k}(\eta)\ .
\ee

The Hamiltonian in the new variables can be rewritten as:
\be\label{hamnew'}
\hH={1\over 2}p_{\eta}^2+{1\over
2} (p_{\phi}+A_{\phi}^s)^2-{1\over 8}\er^{-2\eta}\ , \qq
A_{\phi}^s=\mu \er^{-\eta}\ ,
\ee
in terms of the self-adjoint first order operators
\be\label{saoper}
p_{\phi}=-\i\er^{-\eta}\partial_{\phi}\ ,\qq
p_{\eta}=-\i\er^{-\eta}(\partial_{\eta}+1/2)\ .
\ee

The \nc version is obtained via quantization of the Poisson structure
(\ref{Poissonnew}). The operator
realization of (\ref{Poissonnew}), determined by the commutation
relations:
\be\label{crnew}
[\hat{\eta},\hat{U}]=\xi\hat{U}\ , \qq\hat{\eta}=-\i\xi\partial_{\phi}\ ,\qq
\hat{U}=\er^{\i\phi}\ ,
\ee
can be achieved in an auxiliary Hilbert space $\cF=L^{2}(S^1,d\phi)$; the dimensionless
parameter $\xi$ is
related to the parameters $\theta$ and $\rho$ by $\xi=\theta/\rho^2$. The derivative
$\partial_{\phi}$ is specified by assuming
$\hat{\eta}f_n(\phi)=\xi n f_n(\phi)$, where
$f_n(\phi)=(2\pi)^{-1/2}\er^{\i n\phi},\ n\in {\Z}$.

The wave functions will be operators of the form
\be\label{opform}
\Psi(\hat{\eta},\hat{U})=\sum_{k\in{\Z}}
a_k(\hat{\eta})\hat{U}^k\ ,
\ee
acting in $\cF$ as:
$\Psi(\hat{\eta},\hat{U})f_n(\phi)=\sum_k a_k(n\xi)f_{n+k}(\phi)$.
For the operators (\ref{opform}), we define the
scalar product as follows:
\be\label{scpr'}
(\Psi,\Phi)=2\pi\xi \tr[\er^{2\hat{\eta}}\bar{\Psi}\Phi]\ ,
\ee
where the $\tr$ and the bar are the trace and the hermitian conjugation in $\cF$,
respectively. Inserting here the expansion (\ref{opform}) for $\Psi$ and analogously
for $\Phi$, with $b_k(\hat{\eta})$, we obtain:
\be \label{scpr''}
(\Psi,\Phi)=2\pi\xi
\sum_{k} \sum_{n} \er^{2\xi n}\bar{a}_{k}(n\xi)b_{k}(n\xi)\ .
\ee
Thus, in the \nc case, the integral is replaced by its Riemann sum.

Our next task is to define a \nc analog of the
Hamiltonian (\ref{hamnew'}), i.e., we need \nc analogs
$\hat{p}_{\phi}$ and $\hat{p}_{\eta}$ of the operators (\ref{saoper}). The operators
$\hat{p}_{\phi}$ and $\hat{p}_{\eta}$ are defined by
\begin{eqnarray}\label{delta}
\hat{p}_{\phi}\Psi&=&-\i\hat{\partial}_{\phi}\Psi(\hat{\eta},\hat{U})={1\over
\xi}[\hat{\eta},\Psi(\hat{\eta},\er^{\i\phi})]
=-\i(\partial_{\phi}\Psi)(\hat{\eta},\er^{\i\phi})\ ,\cr
\hat{p}_{\eta}\Psi&=&\frac{\i}{4\sinh(\xi/2)}
[\er^{-\xi/2}\hat{U}\Psi(\hat{\eta},\hat{U})\hat{\bar{U}}-\er^{\xi/2}\hat{\bar{U}}
\Psi(\hat{\eta},\hat{U})\hat{U}]\nbr
&=&\frac{\i}{4\sinh(\xi/2)}[\er^{-\xi/2}\Psi(\hat{\eta}-\xi,\hat{U})-\er^{\xi/2}\Psi(\hat{\eta}+\xi,\hat{U})]\ .
\end{eqnarray}
It is readily seen that the operator $\hat{p}_{\phi}$ 
is self-adjoint with respect to (\ref{scpr''}). As for the operator
$\hat p_\eta$, it is a first order difference operator (in accordance with the spectrum
of $\hat\eta$) such that $\hat p_\eta$ is self-adjoint with respect to (\ref{scpr''}). 
Moreover, in the limit $\xi\rightarrow 0$, we obtain
$\hat{p}_{\eta}\rightarrow -\i(\partial_{\eta}+{1\over 2})$.

The \nc analog of the Hamiltonian is then given as
\be\label{ncham}
\hH\Psi={1\over 2}\hat{p}_{\eta}^2\Psi
+{1\over 2}(\hat{p}_{\phi}+\hat{A}_{\phi}^s)^2\Psi-{1\over 8}\Psi
\er^{-2\hat{\eta}}\ ,
\ee
with $\hat{A}_{\phi}^s\Psi=\mu\Psi\er^{-\eta}$
representing the solenoid magnetic field. Explicitly,
\be\label{hamilt}
\hH\Psi(\hat{\eta})=-{1\over 2}\frac{\Psi(\hat{\eta}+2\xi)
-2\Psi(\hat{\eta})
+\Psi(\hat{\eta}-2\xi)}{(4\sinh(\xi/2))^2}\er^{-2\hat{\eta}}\nbr
+{1\over 2}(\i\partial_{\phi}+\mu)^2\Psi(\hat{\eta})\er^{-2\hat{\eta}}\ ,
\ee
where we have suppressed the explicit $U=\er^{\i\phi}$ dependence of
$\Psi$. We note that the appearance of the factor $\er^{-2\hat{\eta}}$ on the
right-hand side is essential, as it cancels with the weight
factor $\er^{2\hat{\eta}}$ in the scalar product. Inserting the mode expansion (\ref{opform})
into the Schroedinger equation $\hat{H}\hat{\psi}=E\hat{\psi}$,
we obtain a difference equation for the spectral coefficients $a_k(n\xi)$:
\be
-{1\over 2}\frac{[a_k(n\xi+2\xi)-2a_k(n\xi)+a_k(n\xi-2\xi)]}{(4\sinh(\xi/2))^2}
+{1\over 2}(\mu +k)^2a_k(n\xi)=E a_k (n\xi)\er^{2 n\xi}\ .
\ee
Its solution can be
easily obtained in terms of $q$-Bessel functions, with $q=\er^{\xi}$.

Let us discuss now the issue of gauge invariance. In the
commutative case, the gauge transformations
\begin{eqnarray}
\Psi(x)&\rightarrow &\omega(x)\Psi(x)\ ,\qq \bar{\Psi}(x)\rightarrow
\bar{\Psi}(x)\bar{\omega}(x)\nbr
A_j(x)&\rightarrow&A_j(x)-\i\omega(x)\partial_j\bar{\omega}(x)
\end{eqnarray}
are generated by an $x$-dependent phase factor
$\omega(x)=\bar{\omega}^{-1}(x)$. This leads to the covariance of a momentum operator
\be
p_j=-\i\partial_j+A_j\rightarrow\omega(x)(-\i\partial_j+A_j)
\bar{\omega}(x)=\omega(x)p_j\bar{\omega}(x)\ ,
\ee
and guarantees the gauge invariance of the Hamiltonian
matrix elements. On ${\R}_0^2$, the gauge field consists in general of two pieces:
\be
A_j(x)=A_j^s(x)+A_j^r (x)\ ,
\ee
where the singular
(solenoid) part $A_j^s(x)$ is given in (\ref{ham}) and $A_j^r(x)$
represents a magnetic field regular on the whole plane ${\R}^2$.
Any gauge transformation on ${\R}_0^2$ is a composition of a
singular gauge transformation $\omega_{\kappa}(x)=\er^{2\pi
\i\kappa\phi}, \kappa\in{\Z}$ and a regular one,
$\omega_{\lambda}(x)=\er^{{\i}\lambda(x)}$, with
$\lambda(x)=\bar{\lambda}(x)$ regular on ${\R}^2$. The singular
gauge transformation shifts the solenoid magnetic flux from $\mu$
to $\mu+\kappa$, whereas $\omega_{\lambda}(x)$ changes $A_j^r(x)$
to $A_j^r(x)+\partial_j\lambda(x)$.

These topological aspects of the gauge transformations should
not be violated in the \nc case. The gauge transformation of
the wave function $\Psi(\hat{x})$ reads:
\be\label{gaugetr}
\Psi(\hat{x})\rightarrow\omega(\hat{x})\Psi(\hat{x})\ ,\qq
\bar{\Psi}(\hat{x})\rightarrow
\bar{\Psi}(\hat{x})\bar{\omega}(\hat{x})\ ,
\ee
where $\omega(\hat{x})=\bar{\omega}^{-1}(\hat{x})$ is a unitary
operator in $\cal H$. We stress that now the order of factors in
(\ref{gaugetr}) is important. Let us now put
\be
\hat{P}_j=\hat{p}_j+A_j(\hat{x})\ ,\qq
\hat{p}_j=(\hat{p}_{\eta},\hat{p}_{\phi})\ ,
\ee
with $\hat{p}_{\eta}$ and $\hat{p}_{\phi}$ given in (\ref{delta}).
The covariance of $\hat{P}_j$,
$\hat{P}_j\rightarrow\omega(\hat{x})\hat{P}_j\bar{\omega}(\hat{x})$
is guaranteed provided the gauge field $A_j(\hat{x})$ transforms
as:
\be\label{potgaugetr}
A_j(\hat{x})\rightarrow\omega(\hat{x})A_j(\hat{x})
\bar{\omega}(\hat{x})-\omega(\hat{x})[\hat{p}_j,\bar{\omega}(\hat{x})]\ .
\ee
We point out that $A_j(\hat{x})=a_j(\hat{x})\er^{-\hat{\eta}}$
is an operator multiplying $\Psi(\hat{x})$ from the left by
$a_j(\hat{x})$ and simultaneously from the right by
$\er^{-\hat{\eta}}$:
\be\label{defa-j}
A_j(\hat{x})\Psi(\hat{x})=a_j(\hat{x})\Psi(\hat{x})\er^{-\hat{\eta}}\ .
\ee
Again, ordering matters, and only with this ordering,
(\ref{defa-j}) represents a self-adjoint operator. The gauge potential $A_j(\hat{x})$
can be separated into a
regular part $A_j^r(\hat{x})$ and the singular solenoid part given by:
\be\label{solfield}
A_{\eta}^s(\hat{x})\Psi(\hat{x})=0\ ,\qq
A_{\phi}^s(\hat{x})\Psi(\hat{x})=\mu \Psi(\hat{x})e^{-2\hat{\eta}}\ .
\ee
The transformations are
again a composition of a regular gauge transformation
$\omega_{\lambda}(\hat{x})$, changing $A_j^r(\hat{x})$, and a
singular one
$\omega_{\kappa}(\hat{x})=\er^{2\pi \i\kappa\phi},\ \kappa\in{\Z}$.
Using eqs. (\ref{potgaugetr})-(\ref{solfield}), it follows straightforwardly that under
$\omega_{\kappa}(\hat{x})$ the solenoid magnetic flux $\mu$ changes to
$\mu+\kappa$. The gauge invariance of all Hamiltonian matrix
elements is obviously guaranteed. We would like to stress the {\it non}-Abelian form of the
transformation law (\ref{potgaugetr}). Consequently, the field
strength $B(\hat{x})$ is given by a non-Abelian formula, too:
\be
B(\hat{x})=\epsilon_{ij}([\hat{p}_i,A_j(\hat{x})]
+A_i(\hat{x})A_j(\hat{x}))\ .
\ee
It can be easily seen that the field
strength transforms covariantly:
$B(\hat{x})\rightarrow\omega(\hat{x})B(\hat{x})\bar{\omega}(\hat{x})$.

\newsection{Casimir energy on a noncommutative cylinder}

In this section we calculate the vacuum energy in the free scalar field theory
in a (2+1)-dimensional noncommutative space-time with two noncommutative space
coordinates and commutative time, one coordinate being compactified on a
circle. In other words, we consider field theories on a noncommutative space-like
cylinder and with commutative time. As we have already mentioned, the commutativity
of time guarantees that the unitarity condition in
NC-QFT is satisfied \cite{GoM}.

\subsection{Field theory on a noncommutative cylinder \e{ftnc}}

The points on a commutative cylinder $C=\{(\phi,x)\in[0,2\pi]\times{\R};\ \phi=0$ and $\phi=2\pi$
identified$\}=S^1\times{\R}$ can be specified through a
real parameter $x\in{\R}$ and two complex parameters $x_\pm
=\rho \er^{\pm\i{\phi}}$, where $\rho$ is basically the cylinder radius and the $x_\pm$ in the
commutative case are not independent and both correspond to one coordinate, $\phi$. The fields on
$C\times\R$ possess the following expansion: 
\be
\Phi (t,x,{\phi})=\sum_{k=-\infty}^{\infty} \intf
\frac{d\omega\,dp }{(2\pi)^2 }{\tilde \Phi} (\omega,k,p)
\er^{\i(px+k{\phi}-\omega t)}  \ .            \e{cyl.2.1}
\ee

On a cylinder,
the corresponding commutator relations (\ref{comrel}) can not be
realized in terms of self-adjoint operators on $L^2(S^1,d\phi)$.
It is known that the problem has to be formulated not via a
Heisenberg-like relation (\ref{comrel}), but in the corresponding
exponential Weyl form, or as $E(2)$ Lie algebra relations. Thus,
in the noncommutative case, the parameters $x,x_\pm$ are replaced (see also \cite{{CDP1},{CDP3}})
by the operators ${\hat x},{\hat x}_\pm$, satisfying the commutation
relations
\be
[{\hat x},{\hat x}_\pm ]=\pm\ld {\hat x}_\pm \ ,\qq
[{\hat x}_+ ,{\hat x}_- ]\ =\ 0\ ,                            \e{cyl.2.2}
\ee
and the same constraint equation as in the commutative
case: ${\hat x}_+ {\hat x}_- =\rho^2$. The dimensionful (with the
dimension of length) parameter $\ld$ is an analog of the
tensor $\th$ in the case of the Heisenberg-like commutation relation in the
flat Minkowski space.
However, in the present case,
the actual parameter of the noncommutativity is the dimensionless
parameter $\xi =\lambda /\rho$.

In analogy with the commutative case, we take the fields
to be operators in ${\cal H}=L^2 (S^1 ,d{\phi})$,
with the operator Fourier expansion as
\be
\Phi (t,{\hat x},{\hat {\phi}})=\sum_{k=-\infty}^\infty
\int_{-\pi/\lambda}^{+\pi /\lambda} \frac{dp }{2\pi }
\intf\frac{d\omega}{2\pi}\,{\tilde \Phi}
(\omega,k,p ) \er^{\i(p\hat{x}+k{\hat{\phi}})-\i\omega{t}}\ .       \e{cyl.2.4}
\ee
For simplicity, we consider a real scalar field theory which corresponds
to the condition $\Phi^\da (t,{\hat x},{\hat {\phi}})=\Phi (t,{\hat x},{\hat{\phi}})$.
It is important to note that since the spectrum of ${\hat x}$ is discrete:
$x=\lambda n$, $n\in{\Z}$, the integration over $dp$ goes only over a
finite interval $[-\pi /\lambda ,+\pi /\lambda ]$.  We point out that the
operator Fourier expansion \r{cyl.2.4} is invertible:
\be
{\tilde \Phi}(\omega,k,p)=\intf\,dt\,\frac{1}{2\pi }{\tr}\Big[
\er^{-\i k{\hat{\phi}}+\i\omega{t}} \Phi (t,{\hat x},{\hat
{\phi}})\Big]\ .                                         \e{cyl.2.5}
\ee
Then, straightforwardly from the above formula, it follows that
\be
\frac{1}{2\pi }{\tr}[ \er^{-\i k'{\hat{\phi}}-\i p'{\hat x}}
\er^{\i k{\hat{\phi}}+\i p{\hat x}} ]=\delta_{k'k} \delta^{(S)}
(\lambda p' -\lambda p )\ ,                        \e{cyl.2.6}
\ee
where $\delta^{(S)}(\vf)$ denotes the $\delta$-function on a circle.
The {\it usual} inverse Fourier transform of ${\tilde \Phi} (\omega,k,p)$
yields an analog of the Weyl symbol $\Phi(t,n\ld,{\phi})$ on the cylinder:
\be
\Phi (t,n\ld,{\phi})=\intf\,dt\,\sum_{k=-\infty}^\infty
\int_{-\pi/\lambda}^{+\pi /\lambda} \frac{d p }{2\pi } {\tilde \Phi}
(\omega,k,p ) \er^{\i( k{{\phi}}+\ld np-\om t)}\ .       \e{cyl.2.7}
\ee

The star-product for the fields $\Phi(t,n\ld,{\phi})$
has a form which is very
close to that appearing in the \nc plane:
\bn
\Phi_1(t,n\ld,{\phi})\star \Phi_2(t,n\ld,{\phi})&=&
\er^{\frac{\i\ld}{2}\left(\pad{\ }{y_1}\pad{\ }{\vf_2}
-\pad{\ }{y_2}\pad{\ }{\vf_1}\right)}\nbr
&\times&\Phi_1(t,n\ld+y_1,{\phi}+\vf_1)
\Phi_2(t,n\ld+y_2,{\phi}+\vf_2)\Bigg|_{\stackrel{\scriptstyle y_1=y_2=0}
{\scriptstyle\vf_1=\vf_2=0}}\ ,\e{cyl.2.8}
\en
where $y_1,y_2,\vf_1,\vf_2$ are auxiliary continuous variables.

The free action for the scalar fields on the noncommutative space $C_{(NC)}\times\R$
has the following form (cf. \cite{CDP1,CDP3}):
\bn
S_0^{(NCcyl)}[\hat\Phi]&=&\pi\ld\ro\tr\intf\,dt\,
\Big\{\big(\pa_t\hat\Phi\big)^2+\frac{1}{\ld^2}[\Hx_+,\hat\Phi][\Hx_-,\hat\Phi]
+\frac{1}{\ld^2}[\Hx,\hat\Phi]^2-m^2\hat\Phi^2\Big\}\nbr
&=&\frac{\ld\ro}{2}\sum_{n=-\infty}^\infty\intf\,dt\,
\int_{-\pi}^\pi\,d{\phi}\,\Big[\big(\pa_t\Phi(t,n,{\phi})\big)^2
-\big(\d_x\Phi(t,n,{\phi})\big)^2\nbr
&&-\frac{1}{\ro^2}\big(\pa_{\phi}\Phi(t,n,{\phi})\big)^2
-m^2\Phi^2(t,n,{\phi})\Big]\ .  \e{cyl.2.10}
\en
In the last expression, we have used the Weyl symbols \r{cyl.2.7} and the
lattice derivative
$$
\d_x\Phi(t,n,{\phi})=\frac{1}{\ld}\Big[\Phi(t,n+1,{\phi})
-\Phi(t,n,{\phi})\Big]
$$
(we have simplified the notation for the field:
$\Phi(t,n\ld,{\phi})\ra\Phi(t,n,{\phi})$).
As usual for the Weyl symbol, the $\star$-product disappears from the trace
for a product of any two operators. In the case of a field theory on a
flat space, this leads to the free action which formally looks as the one
on the commutative space. In the case of cylinder, we have the trace
of noncommutativity even in the free action: it reveals itself in the
appearance of the discrete derivatives.
We stress that this is an intrinsic property
of field theories on noncommutative manifolds with {\it compact} space-like
dimensions.

\subsection{Vacuum energy of the quantum scalar fields on $C_{(NC)}\times\R$}

In what follows, for simplicity, we  consider the massless case, $m=0$.

After the second quantization, the fields (the Weyl symbols)
$\Phi(t,n,{\phi})$ can be presented in the usual form:
\be
\Phi(t,n,{\phi})=\ld\sum_{k=-\infty}^\infty
\int_{-\pi/\ld}^{+\pi /\ld} \frac{d p}{2\pi}
\Big[\Ha_{p,k}u_{p,k}(t,n,{\phi})+\Ha_{p,k}^\da
\bar{u}_{p,k}(t,n,{\phi})\Big]\ , \e{1cyl}
\ee
where $\Ha_{p,k}$, $\Ha_{p,k}^\da$ are the creation and annihilation
operators, respectively, and $u_{p,k}(t,n,{\phi})$
are the solutions of the equation
of motion corresponding to the action \r{cyl.2.10}:
\bn
u_{p,k}(t,n,{\phi})&=&\frac{1}{2\pi\sqrt{2\ro\ld\om_{p,k}}}
\er^{\i(\ld pn+k{\phi}-\om_{p,k} t)}\ ,\nbr
\om_{p,k}&=&\sqrt{\frac{k^2}{\ro^2}
+\frac{4}{\ld^2}\sin^2\frac{\ld p}{2}}\ ,       \e{2cyl}
\en
normalized with respect to the Klein-Gordon scalar product:
$$
\bigg(u_{p,k},u_{p',k'}\bigg)\equiv\i\ld\ro\sum_{k=-\infty}^\infty
\int_{-\pi}^{\pi}{d{\phi}}\,\bar{u}_{p,k}(t,n,{\phi})
\stackrel{\leftrightarrow}{\pa_t}u_{p',k'}(t,n,{\phi})=\d^{(S)}(\ld p-\ld
p')\d_{k,k'}\ .
$$
The operators $\Ha_{p,k}$, $\Ha_{p,k}^\da$ satisfy the standard commutation
relations:
$$
[\Ha_{p,k},\Ha_{p,k}^\da]=\d^{(S)}(\ld p-\ld p')\d_{k,k'}\ .
$$

Since the time variable is a commutative one, the energy density in this field
model has the usual form:
\be
T_{tt}=\frac12\left[\big(\pa_t\Phi(t,n,{\phi})\big)^2
+\big(\d_x\Phi(t,n,{\phi})\big)^2+\frac{1}{\ro^2}
\big(\pa_{\phi}\Phi(t,n,{\phi})\big)^2\right]\ .  \e{3cyl}
\ee
The calculation of the vacuum expectation value gives (see \cite{GradshteynR},
formula~2.576.2)
\bn
\bra{0}T_{tt}\ket{0}&=&\frac{1}{2(2\pi)^2\ro}
\sum_{k=-\infty}^\infty\int_{-\pi/\ld}^{\pi/\ld}\frac{dp}{2\pi}\,
\sqrt{\frac{k^2}{\ro^2}+\frac{4}{\ld^2}\sin^2\frac{\ld p}{2}}\nbr
&=&\frac{4}{(2\pi)^3\ro\ld^2}+
\frac{8}{(2\pi)^3\ro\ld^2}\sum_{k=1}^\infty\sqrt{(k\xi/2)^2+1}\,
{\bf E}\left(\frac{1}{\sqrt{(k\xi/2)^2+1}}\right)\ ,  \e{4cyl}
\en
where ${\bf E}(x)$ is the complete elliptic integral, and $\xi={\lambda\over\rho}$. This sum
is obviously
divergent and should be regularized (however, the power of
divergence here is equal to two, while in the case of a
{\it commutative} cylinder it is equal to three).
The regularization can be achieved by introducing the cutoff factors
$\exx{-\ve\sqrt{(k\xi/2)^2+1}}$ (cf. \cite{BirrellD}):
\bn
\sum_{k=1}^\infty y \ \
{\bf E}\left(\frac{1}{y}
\right)
&\ra& \sum_{k=1}^\infty y
{\bf E}\left(\frac{1}{y}
\right)
\er^{-\ve y}
\cr \nbr
&=&-\pad{\ }{\ve}\sum_{k=1}^\infty{\bf E}
\left(\frac{1}{y}
\right)
\er^{-\ve y}\ ,
\en
with
$$y=\sqrt{(k\xi/2)^2+1}\ .$$
For small values of the parameter $\xi\ll 1$, the sum can be well approximated
by the corresponding integral with the correction term of the order
$\ord{\xi^2}$:
\be
\sum_{k=1}^\infty\,\frac{\xi}{2}\,
{\bf E}\left(\frac{1}{\sqrt{(k\xi/2)^2+1}}\right)
\er^{-\ve\sqrt{(k\xi/2)^2+1}}=
\int_0^\infty\,dx\,{\bf E}\left(\frac{1}{\sqrt{x^2+1}}\right)
\er^{-\ve\sqrt{x^2+1}}-\frac{\xi}{2}\er^{-\ve}+\ord{\xi^2}\ .
\ee

Now we notice that ${\bf E}(1/\sqrt{x^2+1})$ is a smooth very
slowly varying function: on the half-line $[0,\infty]$ it monotonically
increases from 1 at $x=0$ to $\pi/2$ at infinity. Thus the integral can be
estimated as follows:
$$
\int_0^\infty\,dx\,{\bf E}\left(\frac{1}{\sqrt{x^2+1}}\right)
\er^{-\ve\sqrt{x^2+1}}=c\int_0^\infty\,dx\,
\er^{-\ve\sqrt{x^2+1}}=cK_1(\ve)\ ,
$$
where $c$ is a factor of the order of unity and $K_1(\ve)$ is the modified Bessel
function (see \cite{GradshteynR}, formula~3.365.2). Finally, expanding the
Bessel function, we obtain the expression for the vacuum expectation value of
the energy:
\be
\bra{0}T_{tt}\ket{0}=-\frac{4}{(2\pi)^3\ro\ld^2}
+\frac{2c}{\pi^3\ld^3}\left[\frac{1}{\ve^2}
-\frac12\ln\frac{\ve}{2}-\frac14-\frac12 {\sl C}+\ord{\ve^2}\right]\ .\e{5cyl}
\ee
It is seen that the second term does not depend on the cylinder radius, so that
the Casimir energy density (i.e., the difference between energy densities in
compactified and decompactified cases) for a  noncommutative cylinder proves
to be the following:
\be\label{CylCas}
\bra{0}T_{tt}\ket{0}-\bra{0}T_{tt}\ket{0}\bigg|_{\ro\ra\infty}=
-\frac{1}{4(\pi)^3\ro\ld^2}\ .
\ee

Thus, the Casimir energy on a noncommutative cylinder depends on the radius much
slower than in the commutative case (where a simple dimensional reasoning gives
$\bra{0}T_{tt}\ket{0}\sim 1/\ro^3$). However, we note that we have performed the above
calculations for the  $\xi={\lambda\over \rho}\ll 1$ case and consequently the Casimir energy
in the \nc case, in the leading order, is much larger than the commutative counter-part.
Also we recall the $\lambda$ dependence of the Casimir energy (\ref{CylCas}). As we see, the $\lambda
\to 0$ limit is not a smooth one which is physically  originating from the
fact that $\hat{x}$ has a discrete spectrum. 

\newsection{Casimir energy on a noncommutative torus}

A noncommutative two-torus is defined by the
operators $\hU_i$ which satisfy \cite{{Co},{CDS}}
\bn\label{NCT}
\hU_i {\hat X}_j \hU_i^{-1}&=& {\hat X}_j+\delta_{ij}2\pi R_j {\bf 1}\ , \;\;\;\; i,j=1,2 \cr
\hU_i \hU_j &=& \er^{2\pi\Theta_{ij}}\hU_j \hU_i\ ,\ \ \ \ \Theta_{ij}=\Theta\epsilon_{ij}\ ,
\en
where ${\hat X}_i$ are the coordinate operators and ${\bf 1}$ is the unity in the algebra
representing these coordinates, while $R_i$ are the corresponding compactification radii.
$\Theta$ is the noncommutativity parameter; of course, one should note that compared to the
\nc plane case, it
represents the noncommutativity in the unit volume of the torus (and hence it is dimensionless).
In the following, we assume both radii to be equal and denote them by $\rho$, so that the volume of
the torus is $(2\pi\rho)^2$. It is more convenient to use the ``dimensionless coordinates''
\be
{\hat \alpha}_i={{\hat X_i}\over \rho} \ .
\ee
Then, one can show that the eqs. (\ref{NCT}) have the solutions 
\be
\hU_i=\er^{i\Theta_{ij}{\hat\alpha}_j}\ ,\ \ \ \
[{\hat \alpha}_i,{\hat \alpha}_j]={2\pi\over\Theta}\epsilon_{ij}\ .
\ee
In general the above solution can be represented by (infinite) matrices.

However, for rational $\Theta$, i.e. $\Theta={M\over N}$ where $M$ and $N$ are mutually
prime integers, this algebra possesses only
finite-dimensional representations \cite{Weyl,Schw}. A field theory on such a torus
proves to be lattice-like \cite{BarsM,AmbMNS} and $M$ defines the number of
windings after which one returns to the same lattice site.
Though this is not essential, we put $M=1$, for simplicity.

Similarly to the case of the noncommutative cylinder, considered in the
previous section, scalar fields on the noncommutative torus are defined via
the operator Fourier transform
\be
{\hat \Phi} (t,{\hat \a},{\hat {\b}})=
\intf\,d\omega\,\sum_{k,p=-\infty}^\infty{\tilde \Phi}
(\omega,k,p ) \hU_1^k\ \hU_2^p \er^{-\i\omega{t}}\ ,       \e{tor2}
\ee
which again is invertible and allows to define the Weyl symbols:
\be
\Phi (t,x,y)=\intf\,d\omega\,\sum_{k,p=-\infty}^\infty
{\tilde \Phi}(\omega,k,p ) \er^{\i(kx+py-\om t)}\ ,   \e{tor3}
\ee
and the corresponding $\star$-product.
We also note that in the rational $\Theta$ case, the arguments of the Weyl symbols
are discrete, more precisely:
$$
\Phi (t,n,m)=\intf\,d\omega\,\sum_{k,p=-\infty}^\infty
{\tilde \Phi}(\omega,k,p ) \er^{\i{2\pi\over N}(kn+pm-\om t)}\ ,\ \  n,m=0,1,..., N-1\ .
$$

Let us now restrict the consideration to the rational case. In the rational case, the free scalar 
massless field action reads
\be\label{NCTac}
S^{(NCT)}[\hat\Phi]=\frac{(2\pi\ro)^2}{2}\tr\,\intf\,dt\,
\left\{\big(\pa_t\hat\Phi\big)^2
-\frac{1}{(2\pi\ro\Theta)^2}\left[\left(\hU_1\hat\Phi\hU_1^{-1}-\hat\Phi\right)^2
+\left(\hU_2\hat\Phi\hU_2^{-1}-\hat\Phi\right)^2 \right]\right\}\ .
\ee
The $\tr$ in the above action is normalized as $\tr{\ \bf 1}={1\over \Theta}=N$.
The operator $\hU_1\hat\Phi\hU_1^{-1}$ shifts the argument $n$ of the Weyl
symbol $\Phi (t,n,m)$ by 1, and similarly $\hU_2\hat\Phi\hU_2^{-1}$ shifts $m$.

{\it Note}: For the irrational case, the $\hU_i$ operators cannot serve as
translation operators, and besides them we should define the usual momentum operators.

In terms of the Weyl symbols, the formula (\ref{NCTac}) for the action reads
\be
S^{(NCT)}[\Phi]=
\frac{(2\pi\ro)^2}{2}\sum_{n,m=0}^{N-1}\intf\,dt\,
\Big\{\big(\pa_t\Phi\big)^2-\frac{1}{(2\pi\ro\Theta)^2}\Big(\d_n\Phi\Big)^2
-\frac{1}{(2\pi\ro\Theta)^2}\Big(\d_m\Phi\Big)^2\Big\}\ ,       \e{tor4}
\ee
where $\d_n\Phi(t,n,m)=\Phi(t,n+1,m)-\Phi(t,n,m)$ and
$\d_m\Phi(t,n,m)=\Phi(t,n,m+1)-\Phi(t,n,m)$. Thus the free action on
$T_{(NC)}^2\times\R$ (noncommutative torus with {\it rational} noncommutativity parameter
and commutative time) is equivalent to the Hamiltonian lattice theory (see, e.g.,
\cite{Kogut}).

After the decomposition of the fields into the creation and annihilation
operators,
\be
\Phi(t,n,m)=\sum_{p,k=0}^{N-1}
\Big[\Ha_{p,k}u_{p,k}(t,n,m)+\Ha_{p,k}^\da
\bar{u}_{p,k}(t,n,m)\Big]\ , \e{tor5}
\ee
where the modes
\bnn
u_{p,k}(t,n,m)&=&\frac{1}{2\pi\ro\sqrt{2\om_{p,k}}}
\er^{\i (2\pi/N)(pn+km)-i\om_{p,k} t}\ ,\nbr
\om_{p,k}&=&\frac{N}{\pi\ro}\sqrt{\sin^2\frac{\pi p}{N}
+\sin^2\frac{\pi k}{N}}\                                   \e{tor6a}
\enn
are orthonormal with respect to the scalar product
$$
\big(u_{p,k},u_{p',k'}\big)\equiv\i\left(\frac{2\pi\ro}{N}\right)^2\ \sum_{m,n=0}^{N-1}
\bar{u}_{p,k}(t,n,m)\stackrel{\leftrightarrow}{\pa_t}
u_{p',k'}(t,n,m)=\d_{k,k'}\d_{p,p'}\ ,
$$
we are ready to calculate the vacuum expectation $\bra{0}T_{tt}\ket{0}$
of the energy density. Here, $T_{tt}$ is given by the expression:
$$
T_{tt}=\frac12\Big\{\big(\pa_t\Phi(t,n,m)\big)^2
+\frac{1}{(2\pi\ro\Theta)^2}\Big[\big(\d_n\Phi(t,n,m)\big)^2+
\big(\d_m\Phi(t,n,m)\big)^2\Big]\Big\}\ .
$$

Due to the finiteness of all summations, the expectation value can be easily
found to be the following:
\be\label{casT1}
\bra{0}T_{tt}\ket{0}=\frac{N}{16\pi^3\ro^3}\sum_{k,p=0}^{N-1}
\sqrt{\sin^2\frac{\pi p}{N}+\sin^2\frac{\pi k}{N}}\ .         \e{tor6}
\ee
For any given $N$ (which is, in fact, the inverse of the noncommutativity parameter) the
last sum can be easily evaluated numerically.
Thus, on the noncommutative torus with a rational noncommutativity parameter, the
Casimir energy is finite and can be obtained without any regularization.

For the case of large $N$ (small noncommutativity), eq. (\ref{casT1}) can be
replaced with the corresponding integral:
\be\label{casT2}
\bra{0}T_{tt}\ket{0}= +\frac{N^3}{16\pi^4\rho^3}\int_0^\pi dxdy \sqrt{\sin^2 x
+\sin^2 y}\simeq +\frac{N^3}{16\pi^3\rho^3}.
\ee

Finally we would like to summarize our results on the \nc torus Casimir energy:
\\
1) As we see from (\ref{casT2}), the Casimir energy in the \nc torus case
is {\it cubic} in $N={1\over\Theta}$ $-$ this is the same behaviour as in
 the commutative case with the momentum cut-off equal to $N$. Also we note that the $\Theta\to 0$ limit
(while the volume, and hence $\rho$, is fixed), is not a smooth one. \\
2) The Casimir energy is positive and hence leads to a {\it repulsive force} on the torus.\\
3) As it is expected, its $\rho$ dependence is like ${1\over \rho^3}$.

\newsection{Conclusion}
In this paper, we have investigated various topological and metric
aspects of the $\star$-product approach to the field theories on
\nc spaces. We have compared four simple cases: plane,
cylinder, punctured plane and torus.

Although they differ in topological and metric properties, they
possess the same Poisson bracket and this leads to the same
formal Moyal $\star$-product. We have shown that:

i) In order to take into account various global properties, we
have to choose properly the relevant Darboux pair (elementary
bracket) and restrict the $\star$-algebra to the sub-algebra of
field modes on a particular manifold. In the operator approach, this
is equivalent to a unitary realization in the Hilbert space of
the $\star$-algebra in question.

ii) On a \nc punctured plane, we investigated the Aharonov-Bohm
effect. This space, being topologically equivalent to a cylinder,
possesses a different Riemann metric. Although the proper Darboux
pair is identical to the one on cylinder, still the \nc
generalization of the self-adjoint translation generators (with
respect to the corresponding scalar product) has to be found. We
defined them and, consequently, we gauged the \nc version of the
model. The resulting model is gauge invariant, with the gauge
transformations classified by a (topological) magnetic flux
number: the singular {\it finite} transformations change the flux by an
integer (in suitable units), whereas the regular {\it small} ones do
not influence the flux. These properties, reflecting the
topology, are exactly analogous to those of the commutative case.
Thus, the noncommutativity of the underlying manifold does not
influence the effect in question.
We note that the Aharonov-Bohm phase, besides being important for 2+1 dimensional physics
(e.g. quantum Hall effect), can also shed light on the \nc formulation of the
Wilson loops.

iii) Finally, we would like to stress that on a \nc (punctured)
plane, the notion of trajectory is lost, and there is no way to
describe the change of the wave function along a given path. In
other words, the functions entering the \nc $\star$-algebra $\cal
A $ are Weyl symbols and have no direct physical meaning. However,
the propagator is a well-defined element of $\cal A\otimes\cal A$
and on a punctured plane possesses all the basic properties,
e.g., it respects the basic topological and metric properties and
the symmetry of the \nc version of the model we started with.

We have shown that the space noncommutativity has an essential impact on the
Casimir energy: in the case of a cylinder, its dependence on the radius becomes
much slower in comparison with the commutative space, while on the torus (with a
rational parameter of noncommutativity) the Casimir energy proves to be finite
without any regularization. Moreover, we have found that in the case of the NC-torus, the Casimir
energy is positive, in contrast to the case of a usual commutative space.
This can be very important in the large extra-dimensional models \cite{Dvali}:
The repulsive Casimir force can compensate for the attractive forces originating from the
Kaluza-Klein modes and hence this can provide a stabilization of the
compactification radius for extra dimensions.

We would like to point out the fact that due to the ``Morita equivalence'' \cite{{Co},{CDS}},
which is an equivalence between gauge bundles (sections of the corresponding $C^*$-algebra) on
various tori, the {\it rational} noncommutativity is naturally related to a non-zero magnetic flux.
Therefore, it is plausible to interpret the Casimir energy we have found
(\ref{casT1}) as the energy stored in this constant magnetic flux (replacing the
noncommutativity).

It is worth noting that in many papers (see, e.g., \cite{GomisMW,Nam})
devoted to the Casimir effect in the field theories on noncommutative torus, one started from  a
noncommutative plane and after writing the action in terms of the Weyl symbols,
compactify the coordinates to a torus. 
In that analysis, an interacting \nc $\phi^3$ in six dimensions is considered.  
Then the Casimir energy appears as the finite contributions to the mass of
particles coming from the non-planar diagrams at one loop level
\cite{GomisMW}.
This corrections have unusual dependence on the radius of the
torus (they are proportional  to the radius) and may, in general, stabilize a
size of the torus, considered as an extra subspace in a six-dimensional
space-time with commutative usual four-dimensional subspace \cite{Nam}. 
Though,  it seems more natural to start the quantization
(noncommutative deformation) directly in the space under consideration. However, 
the approach in the above mentioned papers is, evidently, equivalent to the direct
quantization on torus with {\it irrational} parameter of noncommutativity (in this case all
the representations of the coordinate algebra are infinite-dimensional). 
We notice that this occurs only due to the {\it infinite} summations over (discrete)
momentum variables. On the contrary, the finite summation, in the case of
a rational parameter of noncommutativity, cannot change the convergence
properties of the four-dimensional momentum integrals, and as a result this phenomenon
disappears.

{\bf Acknowledgements}
The financial support of the Academy of Finland under the Project No. 163394
is greatly acknowledged.
A.D.'s work was partially supported by RFBR-00-02-17679 grant
and P.P.'s work by VEGA project 1/7069/20.


\end{document}